\documentclass[aps,twocolumn,preprintnumbers,%
superscriptaddress,floatfix]{revtex4}

\usepackage[dvips]{graphics}
\usepackage{amsmath,amsfonts,bm}
\usepackage{epsfig}

\begin{document}
\newcommand{\be}{\begin{eqnarray}}
\newcommand{\ee}{\end{eqnarray}}
\def\lsim{\mathrel{\rlap{\lower3pt\hbox{\hskip1pt$\sim$}}
     \raise1pt\hbox{$<$}}} %less than or approx. symbol
\def\gsim{\mathrel{\rlap{\lower3pt\hbox{\hskip1pt$\sim$}}
     \raise1pt\hbox{$>$}}} %greater than or approx. symbol
\def\N{${\cal N}\,\,$}
\def\bi{\bibitem}\def\la{\langle}\def\ra{\rangle}
\newcommand\<{\langle}
\renewcommand\>{\rangle}
\renewcommand\d{\partial}
\newcommand\LambdaQCD{\Lambda_{\textrm{QCD}}}
\newcommand\tr{\mathrm{Tr}\,}
\newcommand\+{\dagger}
\newcommand\g{g_5}

\title{NA60 and BR Scaling in Terms of the Vector Manifestation: A Model Approach}
\author {G.E. Brown}
\affiliation { Department of Physics and Astronomy\\ State
University of New York, Stony Brook, NY 11794-3800}
%\author{Chang-Hwan Lee}
%\affiliation{Department of Physics, Pusan National University,
%              Pusan 609-735 and
%              \\
%Asia Pacific Center for Theoretical Physics, POSTECH, Pohang
%790-784, Korea}
\author{Mannque Rho}
\affiliation{ Service de Physique Th\'eorique,
 CEA Saclay, 91191 Gif-sur-Yvette c\'edex, France}
%\date{\today}
\begin{abstract}
It is pointed out that the comparison between the recent NA60
dimuon data and the so-called ``Brown-Rho (BR) scaling" as
presented at QM2005 is $not$ founded on a correct interpretation
of the prediction of BR scaling as formulated in 1991 and
modernized recently and hence the conclusion drawn by both the
experimental and theoretical speakers that ``BR scaling is ruled
out by NA60" is erroneous and should be disregarded. We use a
simplified model description of how the vector manifestation of
hidden local symmetry theory enters into the dilepton production,
relegating more rigorous discussions to a follow-up paper.
\end{abstract}

%\date{\today}

\newcommand\sect[1]{\emph{#1}---}

\maketitle

\begin{figure}[htb]
\centerline{\epsfig{file=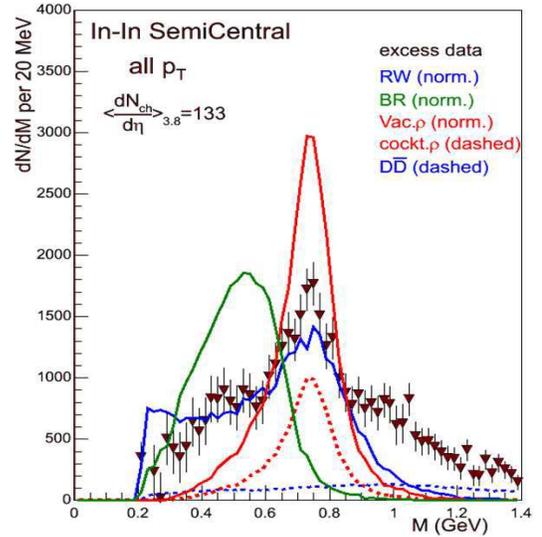,width=7cm,angle=0}} \
 \vskip -0.5cm
\caption{NA60 results compared with the Rapp-Wambach description
(in blue) and so-called ``BR scaling" (in green).} \label{NA60}
 \end{figure}

At Quark Matter 2005, five
talks~\cite{talk1,talk2,talk3,talk4,talk5} flashed Fig.\ref{NA60}
to declare Brown-Rho (BR) scaling ``ruled out" by experiment based
on the ``apparent lack" of that theory as interpreted by
Rapp~\cite{rapp} to explain the NA60 dimuon data. We request that
the readers remove the initials ``BR" from the calculated (green)
curve with our names erroneously associated with it, in that it
has nothing to do with what we believe BR scaling~\cite{BR91} to
be, which is clearly re-outlined recently in \cite{BRDD}. We try
in this note using the simplest possible models and languages, to
explain why the green curve should be removed from the figure, or
at least why our names are not to be associated with it. There are
three main points in our arguments: (1) the ``parametric mass" to
which BR scaling -- hence chiral symmetry aspect -- is directly
connected and the physical (or pole) mass of the vector meson
should be distinguished; (2) BR scaling which has to do with the
``intrinsic property" of the vacuum and Rapp-Wambach (RW)
mechanism~\cite{RW} which has to do with many-body effect
(``sobar" excitations) should be ``fused," because both are
present; (3) Vector dominance (VD), which was employed in
calculating the green curve, is badly violated over most of the
range of temperatures and densities involved and it is obvious
that VD gave most of the shape and height of the green curve.
Schematizing the Harada-Yamawaki renormalization group
approach~\cite{HY} down to bare minimum, we shall show how this
comes about.

We give a more precise and comprehensive discussion in the
accompanying paper~\cite{BR-na60-r} where all three points given
above are treated in light of the modern development of hidden
local symmetry theories that represent low-energy QCD, with a
focus on Harara-Yamawaki hidden local symmetry and its vector
manifestation (VM)~\cite{HY:VM} which is the key point of the
theoretical framework of BR scaling. In this paper, we will focus
on the points (2) and (3) with only a brief reference in a
footnote (see \cite{footnote-b}) to the point (1) which requires a
precise theoretical definition.

Point (2) is easily dealt with because Rapp calculated the fusion
of BR and RW in Fig.5 of \cite{BRDD}. As will be noted below, he
incorrectly used vector dominance (VD) in doing so, but he did
show that the ``fusing" changed both BR and RW curves
substantially. This is because if the $\rho$-meson mass
drops~\cite{footnote-a}, as in BR, the VD coupling $m_\rho^2/g$
should be replaced by ${m_\rho^*}^2/g^*$ (where $g$ stands for the
gauge coupling constant) in the amplitude for dilepton production.
As explained in ~\cite{BR-na60-r}, both the physical $\rho$-meson
mass and the gauge coupling are expected to start dropping when
temperature reaches $\sim 125$ MeV, so the dilepton production
will have the correction factor $(m_\rho^*/m_\rho)^2$ from $\sim
125$ MeV to the critical temperature $\sim 175$ MeV. Also in this
regime, with dropping $\rho$ mass, more decays of the sobar $N^*
(1520)N^{-1}$ will go by way of the $\rho$-meson and fewer into
dileptons.

Point (3) that vector dominance is inapplicable in generic
hadronic system (with the pionic EM form factor in free space
being an exception) was first established by Harada and Yamawaki
in 2001~\cite{HY:VD} and is extensively reviewed in \cite{BRDD}. A
parameter $a$ in hidden local theory of Harada and Yamawaki which
figures importantly in the EM current (see \cite{BR-na60-r}) is
equal to 2 for  vector dominance, but moves in medium quickly
towards 1, its value at the fixed point -- called vector
manifestation fixed point~\cite{HY:VM} -- that is reached when
chiral symmetry is restored where both $m_\rho^*$ and $g^*$ also
have their fixed point values of zero. The renormalization group
flow of $a$ as function of temperature and/or density is not well
known except at near zero temperature and density and in the
vicinity of the vector manifestation fixed point. Furthermore in
heavy-ion processes, temperature and density must be correlated.
Here we will ignore this correlation. We believe the conclusion we
will arrive at is sound.

Let us take the temperature dependence in the form of the most
naive scaling,
 \be
a(T)=\frac{2}{1+T/T_c}
 \ee
which interpolates between $a=2$ at $T=0$ and $a=1$ at $T=T_c$.
Since the dilepton production is quadratic in $a$, the ratio of
the one with $a$ moving towards its fixed point 1 compared with
the vector dominance one is
 \be
{\int_0^{T_c}\left(\frac{1}{1+T/T_c}\right)^2\frac{dT}{T_c}}\left[
{\int_0^{T_c}\frac{dT}{T_c}}\right]^{-1} =1/2.\label{f1}
 \ee
This means that in our simplest model, the cross section is cut
down by a factor of 2.

We expect the factor of 2 to be the lower limit for two related
reasons:(1) temperature dependence of the vector meson mass; (2)
$a\rightarrow 1$ in baryonic systems.

We have argued in a number of
papers~\cite{softglue1,softglue2,BLR05,BLR-star} that the
``melting" of the soft glue is responsible for hadronic masses to
decrease with temperature. In Fig.1 of Brown, Lee and
Rho~\cite{BLR05}, it is shown from unquenched lattice calculations
that the melting of the soft glue begins at $T=125$ MeV; none is
melted before this. In contrast, the temperature dependence in the
formula used by Rapp~\cite{rapp}
 \be
\frac{m_\rho^*}{m_\rho}=(1-0.15n/n_0)(1-(T/T_c)^2)^{0.3}
 \ee
begins lowering the $\rho$-meson mass from $T\sim 0$ and by
$T=125$ MeV
 \be
[1-(T/T_c)^2]^{0.3}=0.81
 \ee
with $T_c=175$ MeV; i.e., the mass in his calculation decreased
19\% at the point at which it begins decreasing in the lattice
calculations. This is contrary to what one expects from the hidden
local symmetry theory with vector
manifestation~\cite{harada-sasaki,footnote-b}.

In our model, we should spread the 50 MeV (=175 MeV- 125 MeV) in
temperature over the interval $M=0.2 - 0.8$ GeV, but since medium
effects begin only at 125 MeV we should take that as the lower
limits in the integrand  in Eq.(\ref{f1}), which now gives $\sim
0.3$ rather than 1/2. We will actually argue below that in dense
medium, the factor will be 1/4.

As mentioned, the vector dominance coupling has the factor
${m_\rho^*}^2/g^*$ in it and the square enters in the cross
section, so this limits further the contribution from high
temperature region in that Rapp's $m_\rho^*$, as that of BR
scaling, goes to zero as $T\rightarrow T_c$.

Rapp (private communication) estimates that with his
parametrization ``the effect of baryon density and temperature in
the dropping mass are comparable" whereas putting together our
above effects we find the temperature dependent effects to be an
order of magnitude smaller than Rapp does. We should therefore
remove almost all of Rapp's contributions to the green curve from
temperature and concentrate on the density dependent effects.

We know from Trnka et al~\cite{omega} that the $\omega$-meson in
Sn has a density dependence consistent with that of BR scaling. We
expect the same density dependence for the $\rho$-meson since the
$\omega$ and $\rho$ are in the same multiplet of $U(2)$. Indeed
Naruki et al~\cite{KEK} find this to be the case in their
experiments. The mass drop observed is somewhat smaller than that
predicted by BR scaling, $\sim 20$\%. As discussed in
\cite{BR-na60-r}, this difference can be explained by dense loop
corrections that appear at higher order than BR scaling. Now in
the presence of density, it is found~\cite{BRDD,DSB-mr} that the
constant $a$ goes precociously to 1 (see the discussion in
\cite{BRDD}). In fact, this is already realized at the level of
one baryon, namely in the EM form factor of the nucleon where
vector dominance is maximally violated with the constant $a$ going
to 1. This suggests that the dilepton production that takes place
in dense matter must be cut down by a factor of 4 compared with
that obtained by Rapp.

%\sect{Introduction}%

\end{document}